\documentclass{elsart}

\usepackage{graphicx}
\usepackage{amssymb}

\usepackage[square,comma,sort&compress]{natbib}

\usepackage{graphicx}
\usepackage{epsfig}
\usepackage{amsfonts,amsmath}
\def\fsec    {fs}                               

\def\mvc     {{\mbox{$\mathrm{MeV/c^2}$}}}

\def\lc      {{\mbox{$\Lambda_{c}^{+}$}}\ }

\def\kii     {{\mbox{$D^{+} \rightarrow K^-\pi^+ \pi^+$}}\ }
\def\kbii    {{\mbox{$D^{-} \rightarrow K^+\pi^- \pi^-$}}\ }

\def\ccd     {{\mbox{$\Xi_{cc}^{+}$}}\ }

\def\lcki    {{\mbox{$\ccd \rightarrow \lc K^-\pi^+ $}}}
\def\dpk     {{\mbox{$\ccd \rightarrow p D^+ K^-$}}}
\def\Back    {1.38}                             
\def\dBack   {0.13}                             
\def\Nsig    {5.62}                             
\def\Signf   {4.8\sigma}                        
\def\Mass    {3518}                           
\def\dMass   {3}                                
\def\amass   {3518.7}                           
\def\damass  {1.7}                                
\def\Pois    {{\mbox{$6.4 \times 10^{-4}$}}}
\begin{document}

\begin{frontmatter}

\title{%
    Confirmation of the Doubly Charmed Baryon $\ccd$(3520) via its Decay to 
    \mbox{$p D^+ K^-$}}
\date{25 August 2005}


The SELEX Collaboration
\author[Aviv]{A.~Ocherashvili\thanksref{trk}},
\author[Aviv]{M.A.~Moinester},
\author[CMU]{J.~Russ},
\author[SLP]{J.~Engelfried\corauthref{cor}},
\corauth[cor]{Corresponding author.}
\ead{jurgen@ifisica.uaslp.mx}
\author[SLP]{I.~Torres},
\author[Iowa]{U.~Akgun},
\author[PNPI]{G.~Alkhazov},
\author[SLP]{J.~Amaro-Reyes},
\author[PNPI]{A.G.~Atamantchouk\thanksref{tra}},
\author[Iowa]{A.S.~Ayan},
\author[ITEP]{M.Y.~Balatz\thanksref{tra}},
\author[PNPI]{N.F.~Bondar},
\author[Fermi]{P.S.~Cooper},
\author[Flint]{L.J.~Dauwe},
% ...
\address[CMU]{Carnegie-Mellon University, Pittsburgh, PA 15213, U.S.A.}
% ...
\address[Fermi]{Fermi National Accelerator Laboratory,
Batavia, IL 60510, U.S.A.}
% ...
\address[ITEP]{Institute of Theoretical and Experimental Physics,
Moscow, Russia}
% ...
\address[PNPI]{Petersburg Nuclear Physics Institute, St. Petersburg, Russia}
\address[Aviv]{Tel Aviv University, 69978 Ramat Aviv, Israel}
\address[SLP]{Universidad Aut\'onoma de San Luis Potos\'{\i},
San Luis Potos\'{\i}, Mexico}
% ...
\address[Iowa]{University of Iowa, Iowa City, IA 52242, U.S.A.}
\address[Flint]{University of Michigan-Flint, Flint, MI 48502, U.S.A.}
% ...
\thanks[tra]{deceased}
\thanks[trb]{Present address: Advanced Mask Technology Center,
Dresden, Germany}
\thanks[trc]{Present address: University of California at Irvine,
Irvine, CA 92697, USA}
\thanks[trd]{Present address: Instituto de F\'{\i}sica da Universidade
Estadual de Campinas, UNICAMP, SP, Brazil}
\thanks[tre]{Present address: Kafkas University, Kars, Turkey}
\thanks[trf]{Present address: Physik-Department, Technische
Universit\"at M\"unchen, 85748 Garching, Germany}
\thanks[trg]{Present address: The Boston Consulting Group, M\"unchen, Germany}
\thanks[trh]{Present address: Lucent Technologies, Naperville, IL}
\thanks[tri]{Present address: Baxter Healthcare, Round Lake IL}
\thanks[trj]{Present address: University of Alabama at Birmingham,
Birmingham, AL 35294}
\thanks[trk]{Present address: Sheba Medical Center, Tel-Hashomer, Israel}
\thanks[trl]{Present address: S\"uleyman Demirel Universitesi, Isparta, Turkey}
\thanks[trm]{Present address: Legal Department, Oracle Corporation,
Redwood Shores, California}
\thanks[trn]{Present address: DOE, Germantown, MD}
\thanks[tro]{Present address: Solidum, Ottawa, Ontario, Canada}
\thanks[trp]{ Present address: Siemens Medizintechnik, Erlangen, Germany}
\thanks[trq]{Present address: Allianz Insurance Group IT, M\"unchen, Germany}

\begin{abstract}
We observe a signal for the doubly charmed baryon 
\ccd in the decay mode
\dpk\ to complement the previous reported decay 
\lcki\ in data from SELEX, the charm hadro-production experiment at 
Fermilab.  In this new decay mode we observe an excess of $\Nsig$ 
  events over a combinatoric background estimated by event mixing to be 
$\Back \pm \dBack$ events.  The mixed background has Gaussian statistics,
giving a signal significance of $\Signf$. The Poisson
probability that a background fluctuation can produce the apparent signal is 
less than $\Pois$.  The observed mass of this state is 3518 $\pm$ 3 \mvc, 
consistent with the published result.  Averaging the two results gives
a mass of $\amass \pm \damass~\mvc$.  The
observation of this new weak decay mode confirms the previous SELEX suggestion
that this state is a double charm baryon.  The relative branching ratio for
these two modes is 0.36 $\pm$ 0.21.
\end{abstract}

\begin{keyword}
Doubly Charmed Baryon

\PACS 
14.20.Lq
\sep
14.40.Lb
\sep
13.30.Eg
\end{keyword}

\end{frontmatter}

\section{Introduction}
In 2002, the SELEX collaboration reported the first observation of a
candidate for a double charmed baryon, decaying as $\lcki $\cite{ccd}.  The
state had a mass of 3519 $\pm$ 2 \mvc, and its observed width was consistent
with experimental resolution, less than 5 $\mvc$.  The final state
contained a charmed hadron, a baryon,
and negative strangeness (\lc and $\rm{K}^-$), consistent with
the Cabibbo-allowed decay of a \ccd configuration. 
In order to confirm the interpretation of this state as a double charm
baryon, it is essential to observe the same state in some other way.
Other experiments with large charm baryon samples, e.g., the FOCUS
and E-791 fixed target charm experiments at Fermilab or the B-factories,
have not confirmed the double charm signal.  This is consistent
with the SELEX results. The report in Ref.~\cite{ccd} emphasized that 
this new state was produced by the baryon beams ($\Sigma^-$, proton) in 
SELEX, but not by the $\pi^-$ beam.  It also noted that the apparent lifetime
of the state was significantly shorter than that of the $\Lambda_c^+$, which
was not expected in a model calculation based on Heavy Quark Effective 
Theory~\cite{life}.

    Another way to confirm the \ccd is to observe
it in a different decay mode that also involves a final state with baryon
number and charm (not anti-charm).
One such mode, involving only stable charged
particles, is the channel \dpk. Observing a 
mass peak near the \lcki\ peak
at 3519 \mvc reported in Ref.~\cite{ccd} in a channel combining a proton with 
a $\rm{D^+ K^-}$ pair but not a $\rm{D^- K^+}$ pair would
confirm the existence of 
the \ccd state.  Here we report the 
first observation of \dpk.

\section{Experimental apparatus}
%
%
The SELEX experiment used the Fermilab charged hyperon beam at 600 GeV to
produce charm particles in a set of thin foil targets of Cu or diamond.
The negative
beam composition was about 50\% $\Sigma^-$, 50\% $\pi^-$.
The positive beam was 90\% protons. 
A beam Transition Radiation Detector identified each beam particle
as meson or baryon with zero overlap.  The
three-stage magnetic spectrometer is shown elsewhere~\cite{Thesis,SELEX}.  The
most important features are the high-precision, highly redundant,
vertex detector that provided an average proper time resolution of 20~{\fsec}
for the charm decays, a 10 m long Ring-Imaging Cerenkov (RICH) detector that
separated $\pi$ from K up to 165 GeV/c~\cite{RICH}, and a high-resolution
tracking system that had momentum resolution of ${\sigma}_{p}/p<1\%$ for a
\mbox{150\,GeV/c} proton.

The experiment selected charm candidate events using an online
secondary vertex algorithm.  A scintillator trigger demanded an
inelastic collision with at least four charged tracks in the
interaction scintillators and at least two hits in the positive
particle hodoscope after the second analyzing magnet.  Event selection
in the online filter required full track reconstruction for measured
fast tracks ($p\,{\scriptstyle\gtrsim}\,15\,\mathrm{GeV}/c$).  These
tracks were extrapolated back into the vertex silicon planes and
linked to silicon hits.  The beam track was measured in upstream
silicon detectors.  A three-dimensional vertex fit was then
performed.  An event was written to tape if all the fast tracks in the
event were \emph{inconsistent} with having come from a single primary
vertex.  This filter passed 1/8 of all interaction triggers and had
about $50\%$ efficiency for otherwise accepted charm decays.  The
experiment recorded data from $15.2 \times 10^{9}$ inelastic
interactions and wrote $1 \times 10^{9}$ events to tape using both
positive and negative beams. The sample was $65\%$ $\Sigma^{-}$-induced,
with the balance split roughly equally between $\pi^{-}$ and
protons.

The offline analysis selected single charm events with a topological 
identification
procedure.  Only charged tracks with reconstructed momenta were used.
Tracks which traversed the RICH
($p\,{\scriptstyle\gtrsim}\,22\,\mathrm{GeV}/c$) were identified as
protons or kaons if those hypotheses were more likely than the pion
hypothesis.  All other tracks were assumed to be pions.  The primary
vertex was refit offline using all found tracks.  An event 
was rejected if all tracks were consistent with one primary vertex.  For those
events which were inconsistent with a single primary vertex, secondary 
vertices were formed geometrically
and then tested against a set of charge, RICH identification, and mass
conditions to identify candidates for the different single charm states.
Candidate events were written to a charm data summary file.  Subsequent
analysis began by selecting particular single-charm species from that 
set of events.

\section{ Search Strategy}
\begin{figure}[ht]
\begin{center}
\psfig{figure=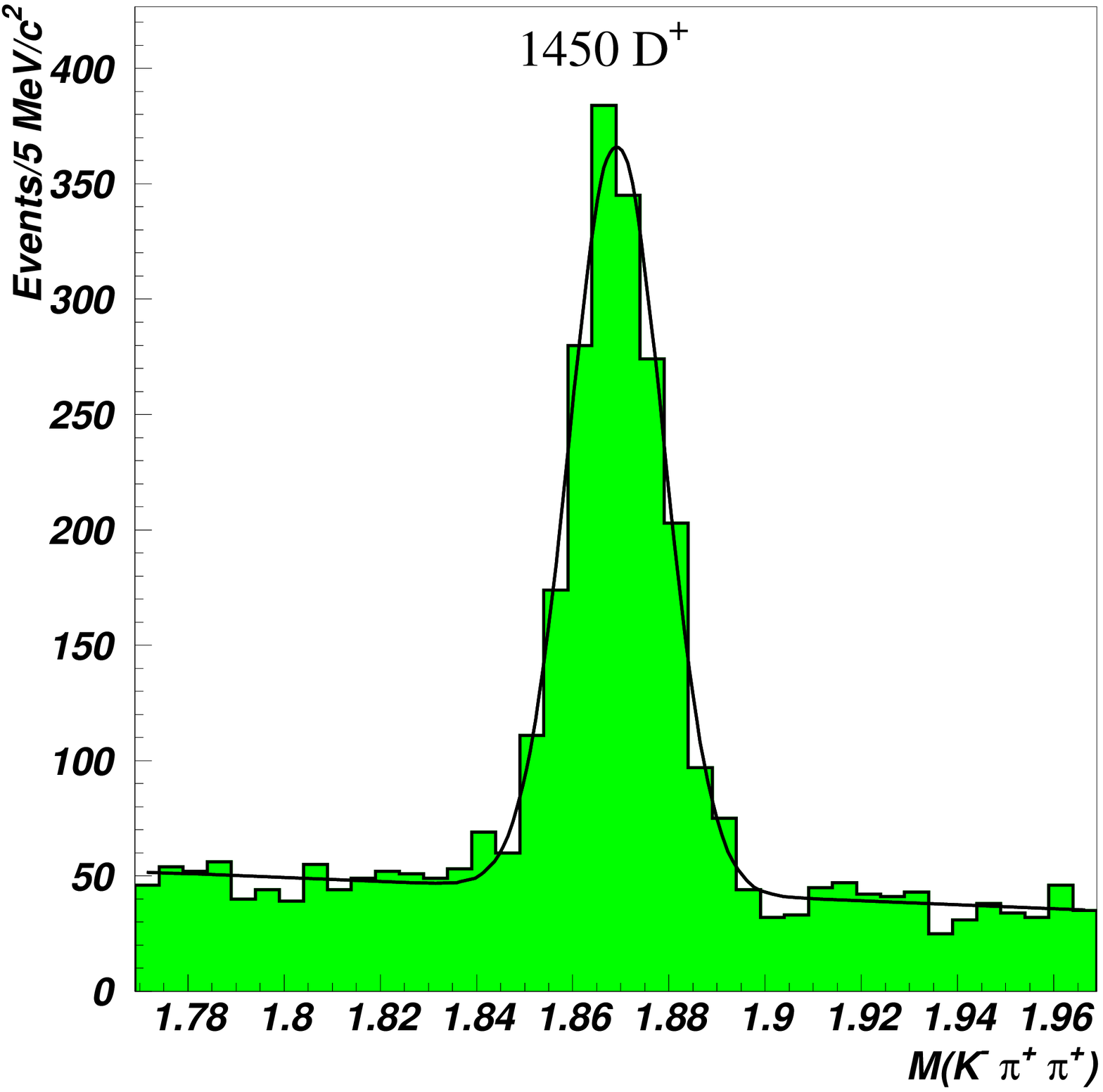,width=67mm} \hspace*{1mm} 
\psfig{figure=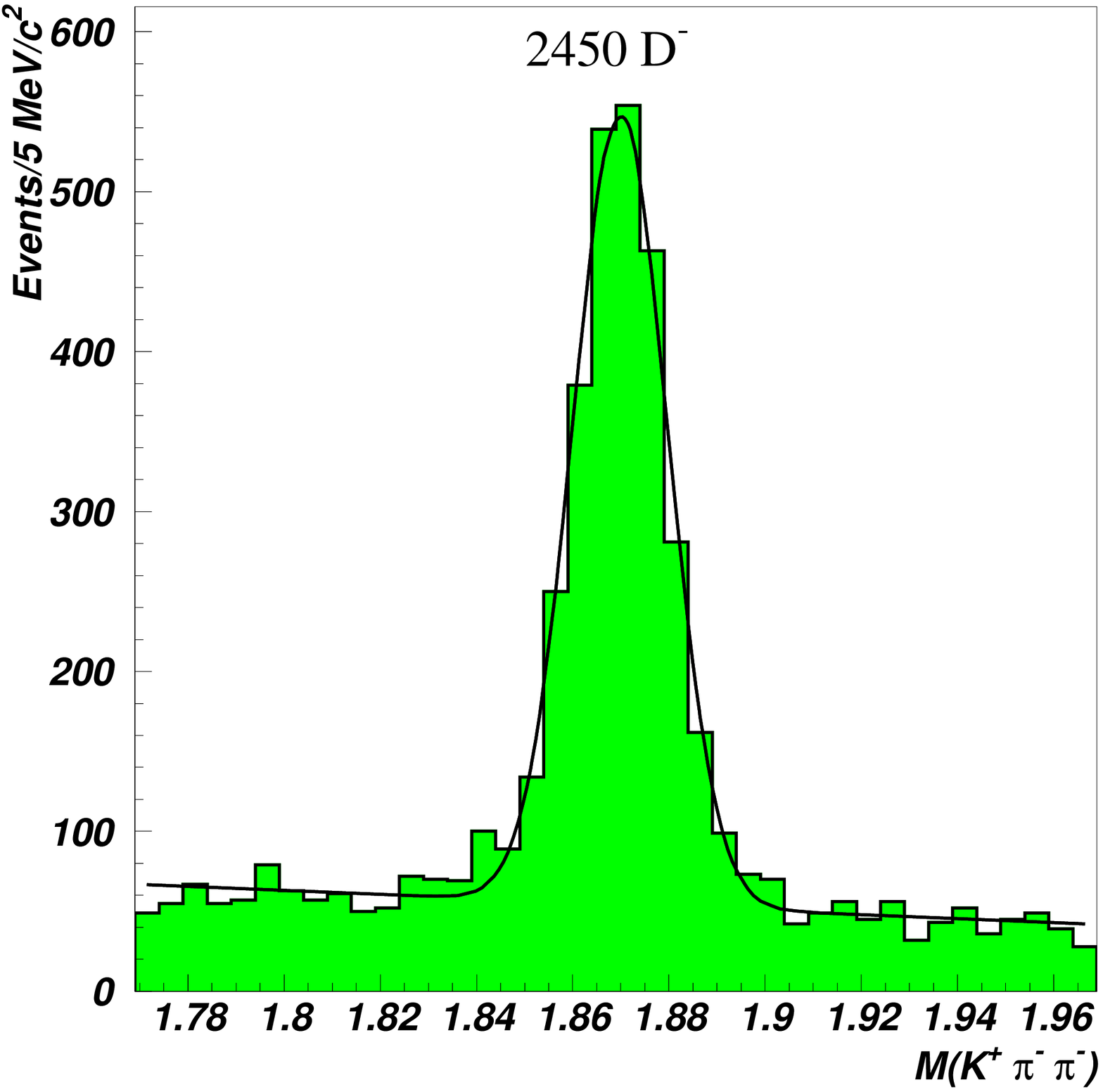,width=67mm}
\caption{\kii(left)~and~\kbii(right) mass distributions with cuts used in 
 this analysis}
\label{dmass}
\end{center}
\end{figure}

In this study we began with the SELEX $\rm{D^{\pm}}$ sample
that has been used in lifetime and hadroproduction studies~\cite{E781-PRL}.
The sample-defining cuts are described in that reference.  No new cuts 
on the D mesons were introduced in this analysis.  The D meson momentum vector
had to point back to the primary vertex with $\chi^2 \ <12$ (the double
charm lifetime is known to be much shorter than the D meson lifetime, so
the D meson pointback is not affected by having come from a secondary 
decay).  The D meson
decay point must have a vertex separation significance of at least 10 $\sigma$
from the primary.  Everywhere in these analyses the vertex error used is the
quadrature sum of the errors on the primary and secondary vertices. The K was 
positively identified by the 
RICH detector.  The pions were required to be RICH-identified if they
went into its acceptance.  The \kii and \kbii mass distributions are
shown in Fig.~\ref{dmass}.  There are 1450~\kii decays and 2450~\kbii decays
in these samples.  The \kii events contribute to the signal channel.  The 
\kbii events cannot come from the decay of a double charm baryon and will be 
used as a topological background control sample. The yield asymmetry stems from
the d-quark contribution of the $\Sigma^-$ beam component that gives a sizeable
production asymmetry favoring leading $\rm{D}^-$ production over $\rm{D}^+$
production, as we have reported for other charm systems~\cite{prod}.

The track-based search code is identical to that used on the 
$\lc$ sample in the original investigation~\cite{ccd}.  The premise is
that a ccd state will make a secondary decay vertex between the primary
production vertex in one of the thin foil targets and the observed D meson
decay vertex, which must lie outside material.  We looked for intermediate 
vertices using all charge
zero pairs of tracks from the set of reconstructed tracks not assigned to 
the D-meson candidate.  
The additional positive track in this final state 
must be RICH-identified as a proton if it traverses the RICH.  The negative 
track in the new vertex is assigned the kaon mass. (This track typically missed
the RICH acceptance by being too soft or too wide-angle,
as confirmed by simulation.)  We made background 
studies by (i) assigning the negative track a pion mass, (ii) looking  
for proton plus positive
track combinations with a D meson, and (iii) looking at proton $\rm{D^- K^-} $
combinations (wrong-sign charm). We require a good 3-prong vertex fit with a 
separation significance of at least 1.0 $\sigma$ from the primary 
vertex, the same requirement used in Ref.~\cite{ccd}. 
The primary position was 
recalculated from the beam track and
secondary tracks assigned to neither the D nor the p$\rm{K}^-$ vertices.  
Results presented in this paper come from this analysis.

\section{Search results and significance}
The signal search was based on a 10 \mvc window centered 
on the \ccd mass 3519 \mvc\ 
from Ref.~\cite{ccd}.
The expected mass resolution for the decay \dpk\ is 4 \mvc.  
Our simulation correctly reproduces the observed widths of all our reported
single charm mesons and baryons.  The 10 \mvc\  window should collect 80\% of
the events in a \ccd signal.
The results are insensitive to changing bin boundaries by up to half a bin.
The background, assumed to be flat, is evaluated outside a 20 \mvc\ window, 
to avoid putting the remaining 20\% of any signal into the background.

\begin{figure}[ht]
\centerline{\psfig{figure=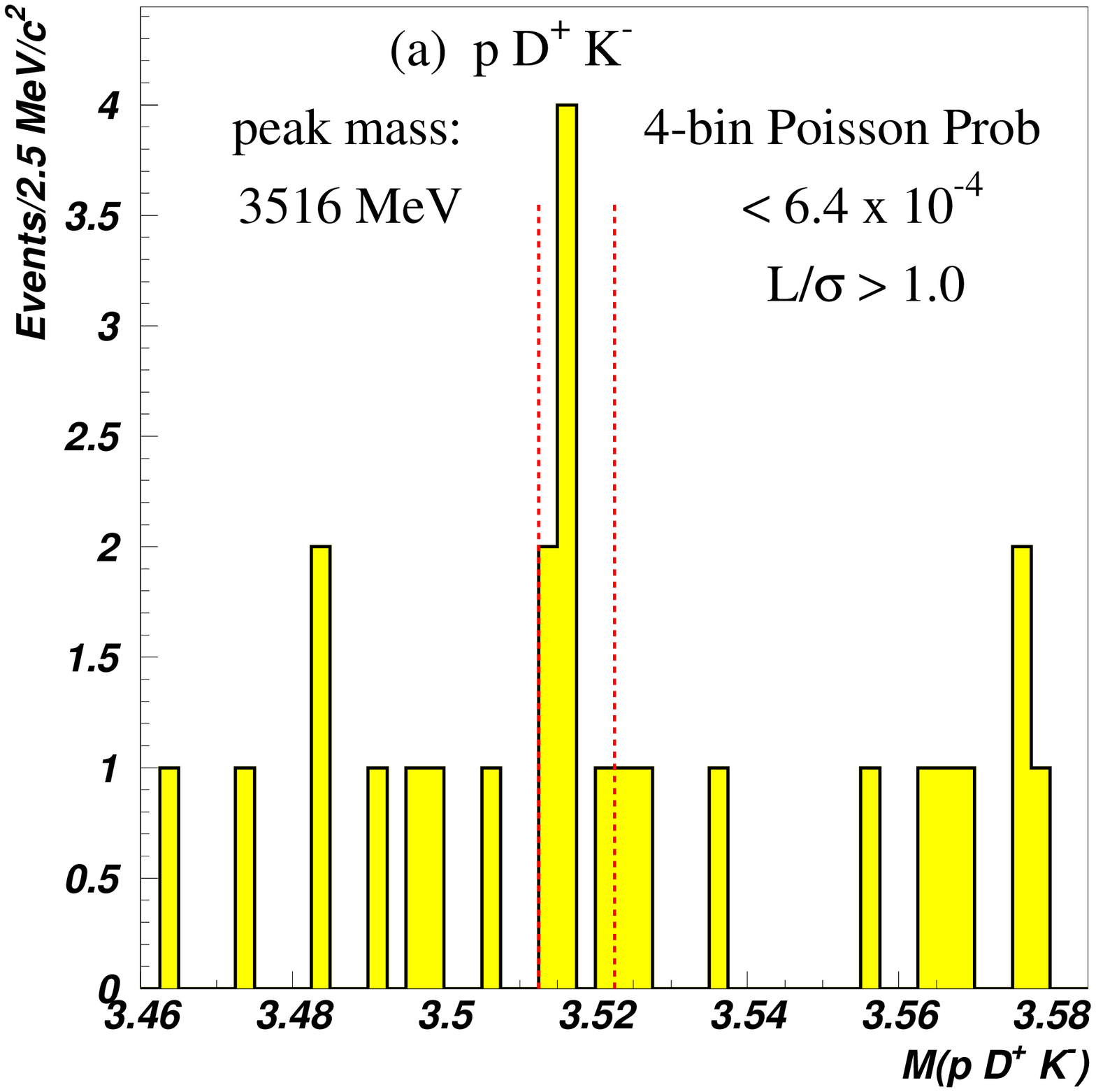,width=90mm}}
\centerline{\psfig{figure=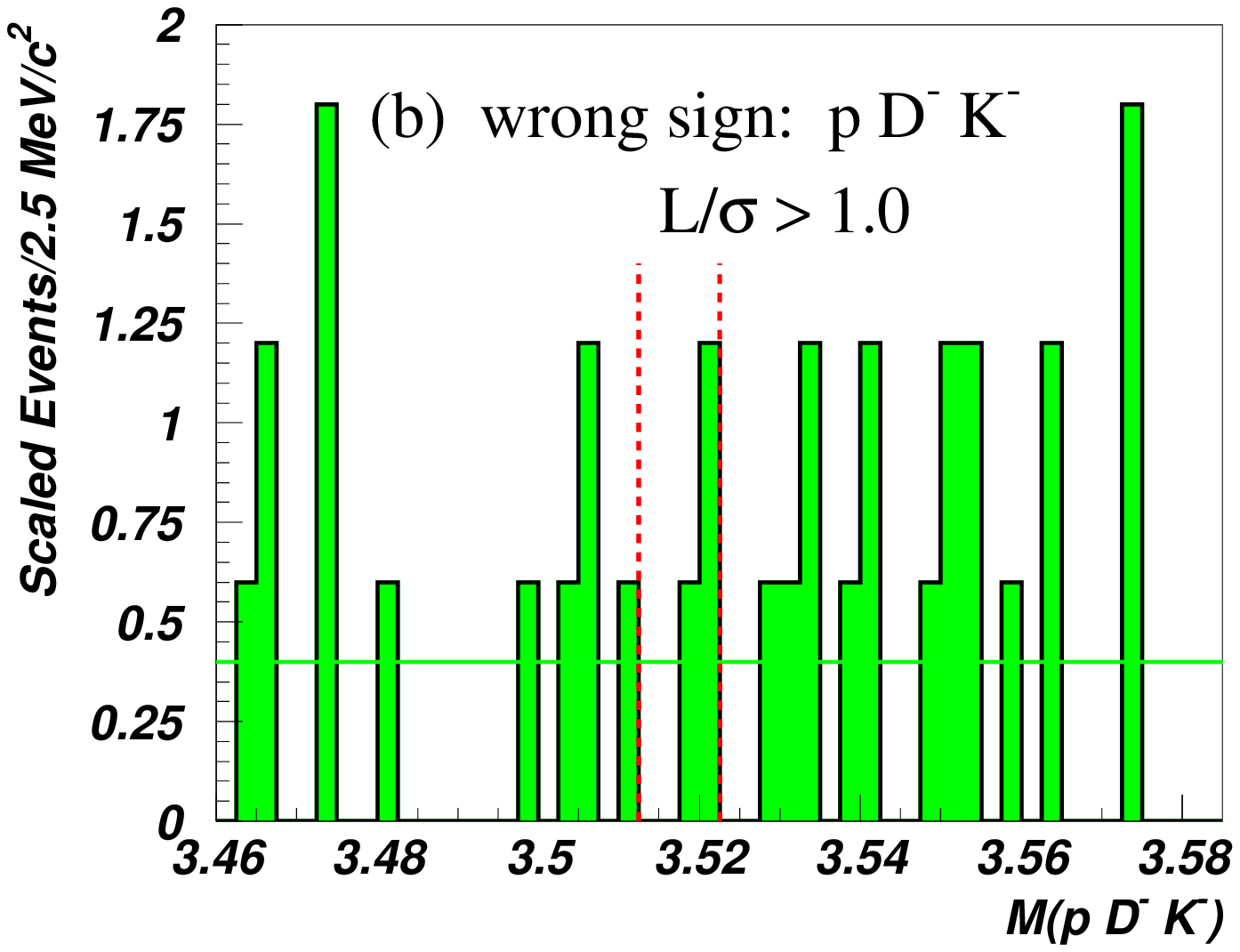,width=90mm}}
\caption{(a) \dpk\ mass distribution for right-sign mass combinations.
Vertical
dashed lines indicate the region of smallest fluctuation probability as
described in the text. (b) wrong-sign 
events with a $\rm{pD^- K^+}$, scaled by 0.6 as described in the text.
The horizontal line shows a maximum likelihood fit to the occupancy.}
\label{dpkmass}
\end{figure}

The right-sign mass combinations in Fig.~\ref{dpkmass}(a) show an excess of
5.4 events over a background of 1.6 events.   The wrong-sign mass combinations 
($ \overline{\rm{c}} $ quark in the decay) 
for the $p\rm{D}^- \pi^+$ final state are also plotted in Fig.~\ref{dpkmass}
(b), scaled by 0.6 for
the $\rm{D^+/D^-}$ratio.  The wrong-sign background shows no evidence for 
a significant narrow structure near 3519 \mvc. 
The average wrong-sign occupancy 
is 0.4 events/bin, exactly the background seen in the right-sign
channel.  This confirms the combinatoric character of the background
population in the right-sign signal.  We have investigated 
all possible permutations of particle assignments.  The only significant 
structure observed is in the channel \dpk, the place where a double charm 
baryon decay can occur.

In order to assign a significance level to this peak, we
have combined statistical
methods used in the original double charm paper~\cite{ccd} and the event-mixing
method used in Ref.~\cite{etaprl}.  We set out to test the hypothesis that the
background events in Fig.~\ref{dpkmass} are random combinatoric
tracks associated
with real $\rm{D^+}$ mesons.  To mix events we took a $\rm{D^+}$ meson in the
peak region (Fig.~\ref{dmass}) and combined it
with proton and $\rm{K^-}$ tracks
extracted from other events.  Each $\rm{D^+}$ was reused 25 times.  To compare 
to the combinatoric background in Fig.~\ref{dpkmass}, we scale the mixed-event 
background down for the multiple $\rm{D^+}$ usage.

\begin{figure}[ht]
\centerline{\psfig{figure=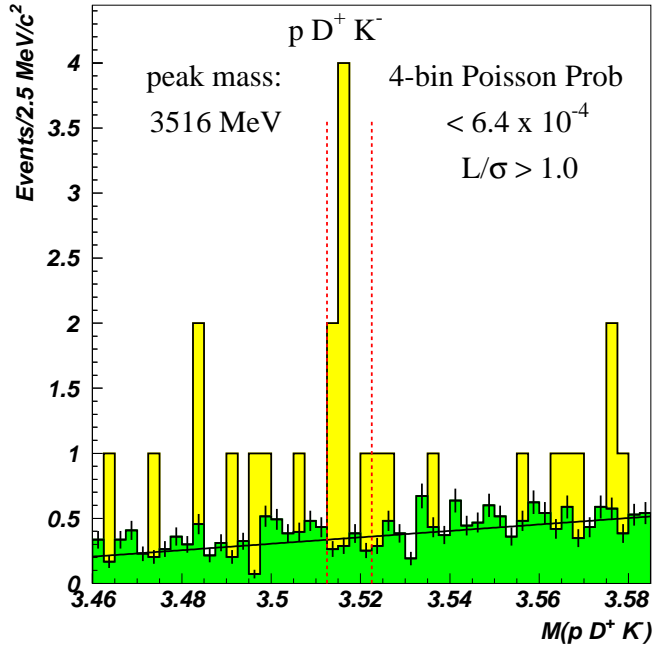,width=90mm}}
\caption{\dpk\ mass distribution from Fig.~\ref{dpkmass}(a)
with high-statistics
measurement of random combinatoric background computed from event-mixing.}
\label{combo}
\end{figure}

The resulting background distribution predicts the observed distribution very
well.
The mean number of background events below, in, and above the 4-bin signal 
peak is 5.74 $\pm$ 0.26, 1.38 $\pm$ 0.13, and 10.60 $\pm$ 0.36. This
agrees well with the 8, 1.6, and 10 events that we observe in the corresponding
regions.  A mass plot with the combinatoric background
level is shown in Fig.~\ref{combo}.  The background that we observe is 
completely consistent in shape and normalization with random combinatoric
tracks associated with real $\rm{D^+}$ mesons.

We take the combinatoric background to give a proper measure of the
expected background under the 7-event signal region.  The single-bin
significance of the signal, using the above background numbers, is:
\quad S = (7 - 1.38)/$\sqrt{1.38 + 0.13^2}$ = 4.8 $\sigma$.  The Poisson
probability of observing at least this much excess, including the Gaussian
uncertainty in the background, is 6.4 x $10^{-4}$.  Both of these
statistical significance calculations use methods identical to those in 
Ref~\cite{ccd}.  This indicates 
a robust signal atop a
combinatoric background whose shape and
normalization are very well understood.  

\section{Signal Properties}
\begin{figure}[ht]
\centerline{\psfig{figure=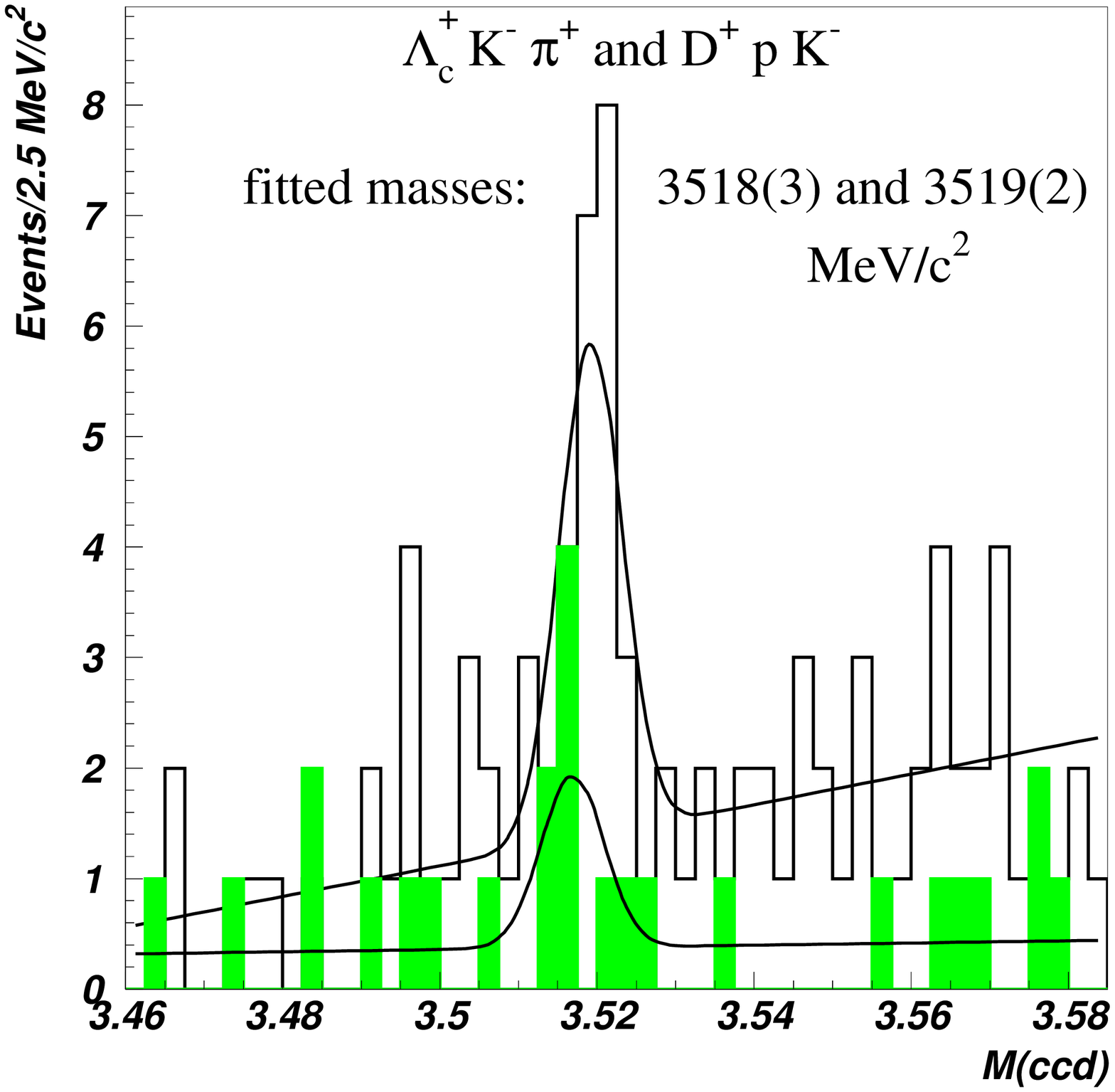,width=90mm}}
\caption{Gaussian fits for \lcki\ and \dpk\ (shaded data) on same plot}
\label{dpkmass_both}
\end{figure}

In order to estimate the mass of the \dpk\ 
state in light of the sparse statistics in Fig.~\ref{dpkmass}, we 
fixed the width of the Gaussian to 4 \mvc and fitted the data distribution 
around the signal peak.
The fit mass is \Mass $\pm$ \dMass \mvc.  This agrees beautifully with
the measurement of 3519 $\pm$ 2 \mvc from the original double charm
baryon report.  We present these data as confirmation of
the double charm state at 3520 $\mvc$ in a new decay mode \dpk. 
The weighted average mass is $\amass \pm \damass$ \mvc.  The mass distributions
for the two channels are shown in Fig.~\ref{dpkmass_both}.

We have used the simulation to study the relative acceptance for
the two decay channels \dpk\ and \lcki\ in order to quote a relative 
branching ratio.  The overall acceptance,
including the single charm selection and the proton ID requirements in the
\dpk\ mode, is very similar.  
SELEX measures the relative branching ratio
$\Gamma$(\dpk)/$\Gamma$(\lcki) $ =  0.36 \pm 0.21$.  The systematic error
due to acceptances is well understood from single charm studies and
is negligible compared to the statistical error.

In Ref.~\cite{ccd} we noted that all observed ccd events were produced by
the baryon beams.  None came from pions.  In this sample, 1 event out of
the 7 in the peak region seen in Fig.~\ref{dpkmass} is a pion beam event,
and 1 of the 19 sideband events comes from the pion beam sample.  This 
sample is consistent with the view that double charm baryons are produced 
dominantly by the baryon beams in SELEX.  In another comparison, we had noted 
that the
\lcki\ decays had an exceptionally short reduced proper time distribution,
indicating a \ccd decay lifetime 5-10 times shorter than the \lc lifetime.
That feature is confirmed by the \dpk\ channel. 
As we noted in Ref.~\cite{ccd},
our lifetime resolution is excellent but we cannot exclude 0 lifetime (strong 
decay) for these events.  The width of this peak is completely consistent
with simulation of a zero-width state, unlikely for a strong decay of a
massive state.  Also, we do not see an increase in the signal when we reduce
the vertex significance cut L/$\sigma$ below 1.  If this were a strong decay, 
one would expect as many events with L/$\sigma$ of -1 as +1, so the signal
should grow significantly.  It does not.

In Ref.~\cite{ccd} we noted that the \lcki\ yield and acceptance implied that
a large fraction of the \lc decays seen in SELEX came from double charm
decays.  That was a surprise.  For the \dpk\ case that is not true.  Only
a few percent of the SELEX $\rm{D}^+$ events are associated with double
charm.

\section{ Summary}

In summary, SELEX reports an independent confirmation of the double charm 
baryon \ccd
previously seen in the \lcki\ decay mode via the observation of its decay
into the \dpk\ final state.  Using only very loose cuts gives
the statistically-significant signal shown in Fig.~\ref{dpkmass}.  
A combinatoric background from event mixing
describes the non-signal distribution
well.  The Gaussian signficance using this background estimate is $\Signf$. 
This decay mode confirms that the \ccd lifetime is very short and that it is
produced dominantly by baryon beams, as we reported in Ref.~\cite{ccd}.

\section*{Acknowledgements}
%
The authors are indebted to the staff of Fermi National Accelerator Laboratory
and for invaluable technical support from the staffs of collaborating
institutions.
This project was supported in part by Bundesministerium f\"ur Bildung, 
Wissenschaft, Forschung und Technologie, Consejo Nacional de 
Ciencia y Tecnolog\'{\i}a {\nobreak (CONACyT)},
Conselho Nacional de Desenvolvimento Cient\'{\i}fico e Tecnol\'ogico,
Fondo de Apoyo a la Investigaci\'on (UASLP),
Funda\c{c}\~ao de Amparo \`a Pesquisa do Estado de S\~ao Paulo (FAPESP),
the Israel Science Foundation founded by the Israel Academy of Sciences and 
Humanities, Istituto Nazionale di Fisica Nucleare (INFN),
the International Science Foundation (ISF),
the National Science Foundation (Phy \#9602178),
NATO (grant CR6.941058-1360/94),
the Russian Academy of Science,
the Russian Ministry of Science and Technology,
the Secretar\'{\i}a de Educaci\'on P\'ublica (Mexico)
(grant number 2003-24-001-026),
the Turkish Scientific and Technological Research Board (T\"{U}B\.ITAK),
and the U.S. Department of Energy (DOE grant DE-FG02-91ER40664 and DOE contract
number DE-AC02-76CHO3000).


\begin{thebibliography}{00}

\bibitem{ccd} M.~Mattson {\sl et al.},
                 Phys.\ Rev.\ Lett.\ {\bf 89}, 112001 (2002).

\bibitem{life} V.V.~Kiselev, A.K.~Likhoded, A.I.~Onishchenko,
Phys.\ Rev.\ {\bf D60} (1999) 014007; Eur.\ Phys.\ J.\ {\bf C16} (2000) 461.\\
B.~Guberina, B.~Melic, H. Stefancic, Eur.\ Phys.\ J.\ {\bf C9} (1999) 213;
{\it ibid}, {\bf C13} (2000) 551.

\bibitem{Thesis} 
  M. Mattson, Ph.D. thesis, Carnegie Mellon University, 2002.

\bibitem{SELEX}  SELEX Collaboration, J.S.~Russ {\sl et al.}, 
                 in {\it Proceedings of the 29th International Conference
                 on High Energy Physics}, 1998, 
                 edited by A. Astbury {\sl et al.} 
                 (World Scientific, Singapore, 1998), Vol. II, p. 1259;
                 hep-ex/9812031.

\bibitem{RICH}   J.~Engelfried {\sl et al.}, 
                 Nucl.\ Instrum.\ Methods A {\bf 431}, 53 (1999).

\bibitem{prod}   F. Garcia {\sl et al.},
                 Phys.\ Lett. B528 (2002) 49.\\
		 M. Kaya {\sl et al.},
		 Phys.\ Lett. B558 (2003) 34.

\bibitem{E781-PRL} A.~Kushnirenko {\sl et al.},
                 Phys.\ Rev.\ Lett.\ {\bf 86}, 5243 (2001), hep-ex/0010014.

\bibitem{etaprl}  A.V.Evdokimov {\sl et al.}, 
                  Phys.\ Rev.\ Lett.\ {\bf 93}, 242001 (2004), hep-ex/0406045

\end{thebibliography}
\end{document}